\documentclass[a4paper,11pt]{article}

\usepackage{iftex}
\ifpTeX
  \PassOptionsToPackage{dvipdfmx}{hyperref}
  \PassOptionsToPackage{dvipdfmx}{geometry}
  \PassOptionsToPackage{dvipdfmx}{xcolor}
  \PassOptionsToPackage{dvipdfmx}{graphicx}
\fi

\usepackage[T1]{fontenc}
\usepackage{fix-cm}

\usepackage{url}
\usepackage{hyperref}
\usepackage{authblk}
\usepackage{geometry}
\usepackage{amsmath,amssymb,amsthm}
\usepackage{array}
\usepackage[linesnumbered,ruled,vlined,noresetcount]{algorithm2e}
\DontPrintSemicolon

\SetCommentSty{mycommfont}
\SetArgSty{textnormal}
\SetKwInput{Input}{Input}
\SetKwInput{Notation}{Notation}
\SetKwInput{Predicates}{Predicates}
\SetKwProg{Fn}{function}{:}{end}
\SetKwInput{Variables}{Variables}
\SetKwInput{Parameter}{Parameter}
\SetKwInput{Note}{Note}
\SetKwInput{Remark}{Remark}
\SetKwInput{Initially}{Initially}

\usepackage{multirow}
\usepackage{xcolor}
\usepackage{graphicx}
\usepackage{xfrac}
\usepackage{tikz}
\usetikzlibrary{arrows.meta}
\usepackage{needspace}

\newtheorem{theorem}{Theorem}
\newtheorem{corollary}{Corollary}
\newtheorem{lemma}{Lemma}

\theoremstyle{definition}
\newtheorem{definition}{Definition}

\newcommand{\mset}[1]{\mathopen{\{\!\lvert}#1\mathclose{\rvert\!\}}}

\newcommand{\R}{\mathbb{R}}

\newcommand{\calA}{\mathcal{A}}

\newcommand{\tomax}{\textsc{ToMax}}

\title{Byzantine Resilience of Robot Gathering under Round-Robin Scheduling}
\date{}

\author[1]{Fabian Frei}
\author[2]{Yuichi Sudo}
\author[2]{Koichi Wada}

\affil[1]{MIT, MA, USA}
\affil[2]{Hosei University, Tokyo, Japan}

\begin{document}

\maketitle

\begin{abstract}
We study deterministic gathering of $n$ anonymous, oblivious
mobile robots in the plane under round-robin scheduling with
rigid movement, in the presence of $f$ Byzantine robots.
Robots observe all robot positions with their multiplicities, while their local coordinate systems may change adversarially between activations.
The goal is for all correct robots to gather at a single point in finite time and remain there forever.
We analyze \tomax, a simple algorithm that moves to a point of maximum multiplicity and stays still if and only if its current position is the unique maximum.
We prove that \tomax{} achieves gathering whenever $f\le\lfloor\sqrt{n}/2\rfloor$, establishing an $\Omega(\sqrt{n})$ lower bound on the achievable Byzantine resilience.
\end{abstract}

\thispagestyle{empty}
\addtocounter{page}{-1}
\clearpage

\section{Introduction}
\label{sec:intro}

We study the Byzantine resilience of deterministic gathering by Look--Compute--Move (LCM) robots~\cite{SY99} under round-robin scheduling.
The gathering problem requires all correct (non-Byzantine) robots to reach a common point in finite time and remain there forever, regardless of the adversarial behavior of the Byzantine robots.
Our focus is on how many Byzantine robots 
can be tolerated as the number of robots grows.

Prior work has established a range of impossibility results for Byzantine gathering~\cite{AP06,DGM+06,IBTW11,DPR20}, while positive guarantees remain scarce.
Agmon and Peleg~\cite{AP06} showed that gathering is possible under fully synchronous scheduling (FSYNCH) whenever $n\ge3f+1$, where $n$ is the total number of robots and $f$ is the number of Byzantine robots.
This permits a constant fraction of the robots to be Byzantine.
D\'efago, Gradinariu, Messika, and Raipin-Parv\'edy~\cite{DGM+06} noted that the maximum-multiplicity strategy achieves gathering under round-robin scheduling (RR) for $n=3$ and $f=1$.
It is straightforward to verify that the same strategy achieves gathering for every $n\ge3$ with $f=1$.
However, to the best of our knowledge, no previous guarantee for deterministic gathering under RR permits the number of Byzantine faults to grow with $n$.

Both FSYNCH and RR impose strong regularity on robot activations.
Under FSYNCH, all robots are activated simultaneously; under RR, they are activated one at a time in a fixed cyclic order.
Despite this regularity in both settings, the available guarantees differ substantially: a linear number of Byzantine faults under FSYNCH, but only a single fault under RR.

This paper advances the resilience guarantee under RR from one Byzantine robot to $\Theta(\sqrt n)$ Byzantine robots.
Specifically, we prove that gathering is possible whenever $n\ge4f^2$, giving the resilience lower bound $\psi(n)=\Omega(\sqrt n)$.
Here, $\psi(n)$ denotes the largest number of Byzantine faults that can be tolerated by a gathering algorithm for $n$ robots.
To the best of our knowledge, this is the first superconstant resilience guarantee for deterministic Byzantine gathering under a scheduler other than FSYNCH.

A substantial gap remains between our sufficient condition $n\ge4f^2$ and the general necessary condition $n>2f$ for tolerating up to $f$ Byzantine robots.
In particular, whether RR can support gathering despite a constant fraction of Byzantine robots, as FSYNCH does, remains open.
Our result provides a starting point for determining how Byzantine resilience scales with the number of robots beyond fully synchronous activation.

\subsection{Related Work}
\label{sec:related-work}

\paragraph{LCM robots and model assumptions.}
The LCM model, introduced by Suzuki and Yamashita~\cite{SY99}, provides a formal framework for studying distributed algorithms for mobile robots.
The model has since been extensively studied, particularly in the context of fundamental coordination problems such as gathering and pattern formation.
In each LCM cycle, a robot observes its surroundings (\emph{Look}), computes a destination (\emph{Compute}), and moves toward it (\emph{Move}).
Variants of the model differ in the robots' capabilities and the assumptions on their environment.
Oblivious robots retain no information between cycles, whereas non-oblivious robots can retain internal state.
In the luminous model~\cite{DFP+12,DFP+16}, each robot has a persistent light whose color is drawn from a finite set and is visible to itself and other robots.

The solvability of various problems, including gathering, depends on the scheduling model, which dictates the timing and order of robot activations.
Under fully synchronous (FSYNCH) and semi-synchronous (SSYNCH) scheduling, execution proceeds in rounds: FSYNCH activates every robot in each round, whereas SSYNCH activates an arbitrary nonempty subset, whose members execute their cycle phases synchronously.
Under asynchronous scheduling (ASYNCH), phases may overlap across robots, and observations may become outdated before they are acted upon. Centralized (or sequential) schedulers allow only one robot to execute a complete cycle at a time; RR further fixes a cyclic activation order.
A $k$-bounded scheduler allows each other robot at most $k$ activations between consecutive activations of any given robot.

The capabilities of robots in terms of movement and sensing provide further distinctions.
With \emph{rigid} movement, a robot always reaches its computed destination; with \emph{non-rigid} movement, a move may be interrupted, subject to a fixed positive minimum progress guarantee, with closer destinations reached.
For sensing, robots may or may not have the ability to detect multiplicity at a location.
Without multiplicity detection, a snapshot reveals only occupied locations.
\emph{Weak multiplicity detection} distinguishes one robot from multiple robots at each occupied location, while \emph{strong multiplicity detection} reveals the exact number.

This paper assumes rigid movement and strong multiplicity detection with RR scheduling, with all robots, including Byzantine robots, subject to the same cyclic activation order.

\paragraph{Byzantine gathering in the plane.}
For non-rigid, oblivious robots, \cite{AP06} proves that no deterministic algorithm can achieve gathering or even convergence under SSYNCH when $n=3$ and $f=1$.
Under FSYNCH, the same work gives gathering algorithms for $n\ge3f+1$, as well as for the special case $n=3$ and $f=1$.
These results establish that fully synchronous activation can support gathering despite a linear number of Byzantine faults.

The maximum-multiplicity strategy of \cite{DGM+06} is the closest predecessor of our algorithm \tomax, presented in Section~\ref{sec:lower-bound}.
The paper notes that, with strong multiplicity detection, this strategy tolerates one Byzantine robot among three robots under RR.
For rigid moves and the tie rule used in \tomax, the extension to every $n\ge3$ with $f=1$ is immediate: between two consecutive activations of the Byzantine robot, all $n-1$ correct robots are activated consecutively.
Consider the $n-1$ consecutive correct activations immediately following an activation of the Byzantine robot.
After the first correct activation, there is a unique maximum-multiplicity point: if the robot moves, its destination gains one robot and becomes uniquely maximal; otherwise, its current position was already uniquely maximal.
Every subsequent correct robot moves to that point or remains there.
Since the Byzantine robot is not activated during this interval, all $n-1$ correct robots gather at that point.
As $n-1>1$, the point remains uniquely maximal regardless of the Byzantine robot's subsequent behavior, so all correct robots remain stationary.

Note~5.2 of~\cite{DGM+06} also claims that the same strategy tolerates one Byzantine robot for every odd $n\ge5$ under any fair centralized scheduler.
However, this claim is false, as shown by the following counterexample with $n=5$.
Place two correct robots $a_1,a_2$ at $p$ and two correct robots $c_1,c_2$ at a distinct point $q$.
Under the fair schedule obtained by repeating
$(b,a_1,a_2,b,c_1,c_2)$ forever, let the Byzantine robot $b$ move to $p$ at its first activation in each period and to $q$ at its second.
Every correct robot is activated at the unique maximum-multiplicity point and therefore remains stationary, so gathering never occurs.

The journal version by D\'efago, Potop-Butucaru, and Raipin-Parv\'edy~\cite{DPR20} establishes only impossibility results for Byzantine gathering; the positive guarantees claimed in \cite{DGM+06} are not retained.
In particular, Lemmas~16--18 rule out deterministic gathering for oblivious robots under centralized $k$-bounded schedulers, even with multiplicity detection, in the following cases:
(i) $n=3$, $f=1$, and $k\ge2$;
(ii) $n$ is even, $f\ge1$, and $k\ge\lceil(n-f)/f\rceil$; and
(iii) $n$ is odd, $f\ge2$, and $k\ge\lceil(n-f)/(f-1)\rceil$.
Without multiplicity detection, Lemma~15 also rules out both deterministic and randomized weak gathering under RR for $f\ge1$ and $n>f+1$.

Izumi, Bouzid, Tixeuil, and Wada~\cite{IBTW11} use a reduction based on BG-simulation to establish another impossibility result for deterministic Byzantine gathering.
Translated into our notation, their result applies to $n\ge4$ robots with one Byzantine robot under an $(n-1)$-bounded centralized scheduler, even with multiplicity detection and internal memory.
Note that this does not imply impossibility under the more restricted RR scheduler.

As reviewed above, most known results on Byzantine gathering by LCM robots are impossibility results.
To the best of our knowledge, prior positive guarantees for deterministic gathering are limited to FSYNCH with $n\ge3f+1$ or $(n,f)=(3,1)$~\cite{AP06}, and RR with $n\ge3$ and $f=1$~\cite{DGM+06}.

\paragraph{Round-robin and sequential scheduling.}
RR and more general sequential schedulers have also been studied for gathering and pattern formation without faults.
Frei and Wada~\cite{FW26} study gathering from distinct initial positions under RR without multiplicity detection.
They show that four or more oblivious robots can gather even with non-rigid movement, whereas three robots require additional self-consistency of their local coordinate systems.
For three robots, maintaining either the local unit of distance or the compass across activations suffices, without requiring agreement between robots.
These results isolate the effects of activation order and coordinate consistency even in the absence of faults.

Flocchini, Navarra, Pattanayak, Piselli, and Santoro~\cite{FNP+26} consider arbitrary pattern formation under general fair sequential schedulers.
For oblivious non-rigid robots, they give algorithms for every target pattern with between two and $n$ distinct points without multiplicity detection, and show that weak multiplicity detection is necessary and sufficient for gathering when $n>2$.
Thus, the distinction between gathering and other patterns remains consequential even when simultaneous activations are excluded.
Navarra and Piselli~\cite{NP25} study RR gathering on anonymous unoriented rings without multiplicity detection, characterize solvability from both arbitrary and distinct initial positions, and give asymptotically optimal algorithms.
Terai, Wada, and Katayama~\cite{TWK23} investigate gathering with lights under several schedulers, including RR, relating scheduling assumptions to the robots' available persistent state.
Together, these works examine how sequential activation interacts with geometry, multiplicity detection, and memory; our focus is its interaction with Byzantine faults.

\paragraph{Other gathering models and objectives.}
Bramas and Tixeuil~\cite{BT15} study wait-free gathering without chirality in the presence of crash faults.
Crash failures permanently stop robots, whereas Byzantine robots may continue moving adversarially, so crash-tolerance guarantees do not directly imply Byzantine tolerance.

Byzantine \emph{convergence}, which only requires the distances between correct robots to tend to zero, admits stronger resilience guarantees than those currently known for RR gathering.
For robots on a line, Bouzid, Potop-Butucaru, and Tixeuil~\cite{BPT10} establish tight thresholds $n>2f$ under FSYNCH and $n>3f$ under bounded asynchronous scheduling.
Without a bound on scheduling, \cite{BPT09a} gives an asynchronous convergence algorithm for $n>5f$ and a matching lower bound for cautious algorithms.
Related convergence impossibilities have also been verified in Coq~\cite{ABC+13}.
Location functions for Byzantine-tolerant swarms support tasks such as convergence, marching, and exploration under FSYNCH with shared coordinates~\cite{ADK+23}.
These results concern objectives or capabilities different from finite-time gathering by oblivious robots without coordinate agreement.

Byzantine gathering has also been studied for mobile agents in graphs.
In this setting, agents typically have labels and memory, communicate when colocated, and move in synchronous rounds.
With a known Byzantine bound $f$, Dieudonn\'e, Pelc, and Peleg~\cite{DPP14} achieve gathering with termination even when Byzantine agents can forge labels, requiring at least $2f+1$ correct agents if the graph size is known and $4f+2$ when no upper bound on it is known.
Bouchard, Dieudonn\'e, and Ducourthial~\cite{BDD16} reduce these requirements to the optimal thresholds of $f+1$ and $f+2$, respectively.
\cite{DPP14} also establishes tight bounds for weakly Byzantine agents, which cannot forge labels: any positive number of correct agents suffices when the graph size is known, whereas $f+2$ are necessary and sufficient when no upper bound on the graph size is known.
Hirose, Nakamura, Ooshita, and Inoue~\cite{HNOI22} study the tradeoff between resilience and gathering time for weakly Byzantine agents.
They give polynomial-time algorithms assuming an upper bound on the graph size and at least $4f^2+8f+4$ correct agents; \cite{HNOI24} reduces this requirement to $7f+7$ while retaining polynomial running time and simultaneous termination.
Hirose, Eguchi, and Sudo~\cite{HES25} study self-stabilizing \emph{perpetual gathering} under weak Byzantine faults: from arbitrary initial states, correct agents must eventually remain together, although they may keep moving together.
These graph-based results concern a different computational model and do not directly yield algorithms for anonymous oblivious LCM robots.

\subsection{Our Contribution}
\label{sec:our-contribution}

We consider deterministic gathering by anonymous, oblivious robots under RR scheduling, assuming rigid movement and strong multiplicity detection.
We prove that our algorithm \tomax{} achieves gathering whenever $f\le\lfloor\sqrt{n}/2\rfloor$, without knowing $f$.
This raises the previously known resilience guarantee from one Byzantine robot to $\Theta(\sqrt{n})$ Byzantine robots and establishes $\psi(n)=\Omega(\sqrt{n})$.

We briefly outline our analysis, which exploits the fixed cyclic activation order.
After its first full activation, each maximal block of consecutive correct robots behaves as a single group.
If such a block contains more than $2f$ robots and moves, it creates a multiplicity advantage that the Byzantine robots cannot erase before all correct robots gather.
Thus, every block that moves infinitely often has size at most $2f$.
A separate potential argument shows that all correct robots that eventually stop moving must share the same final position.
Combining these observations with a counting argument, we show that, when $n\ge4f^2$, any execution in which some correct robots move infinitely often would have a strict majority of robots eventually stationary at a common point.
This majority forces all remaining correct robots to gather there, yielding a contradiction.
Hence, all correct robots eventually stop moving at a common point.

We also give a self-contained proof of the general upper bound $\psi(n)\le\lfloor(n-1)/2\rfloor$ and construct a repeating execution showing that the resilience of \tomax{} is at most $\lfloor(n+1)/3\rfloor$.
A substantial gap remains between the square-root lower bound and these linear upper bounds.
Determining the asymptotic order of $\psi(n)$, and in particular whether RR can tolerate a constant fraction of Byzantine robots, remains open.

The remainder of the paper is organized as follows.
Section~\ref{sec:preliminaries} defines the model and resilience.
Section~\ref{sec:lower-bound} presents \tomax{} and proves the lower bound, while Section~\ref{sec:upper-bound} proves the upper bounds.
Section~\ref{sec:conclusion} concludes the paper.

\section{Preliminaries}
\label{sec:preliminaries}
For integers $a$ and $b$, we write $[a,b]=\{a,a+1,\dots,b\}$.
For a multiset $B$ and an element $b\in B$, let $m_B(b)$ denote the multiplicity of $b$ in $B$, that is, the number of occurrences of $b$ in $B$.

We consider $n$ robots $r_0,r_1,\dots,r_{n-1}$ moving in the Euclidean plane $\R^2$.
The robots are oblivious, identical and anonymous.
Throughout the paper, robot indices are taken modulo $n$ whenever this is clear from the context; for example, $r_t=r_{t\bmod n}$.
Exactly $f\in[0,n-1]$ robots are \emph{Byzantine}, and the remaining $n-f$ robots are \emph{correct}.

We assume a round-robin scheduler: at each time $t=0,1,2,\dots$, only robot $r_t$ is activated and executes the Look--Compute--Move cycle described below.
Let $p_t(r)$ denote the position of robot $r$ immediately before the Look phase at time $t$, expressed in a global coordinate system used only for the analysis.
The robots have no direct access to this global coordinate system.
We adopt the \emph{variable disorientation} model: the robots do not share a coordinate system, and even the local coordinate system of an individual robot may change adversarially between activations through rotation, reflection, and uniform scaling.
The local coordinate system used by $r_t$ at time $t$ is represented by a matrix $L_t\in\R^{2\times2}$ satisfying
\[
  L_t=s_tQ_t, \qquad Q_t^{\mathsf T}Q_t=I_2,
\]
for some $Q_t\in\R^{2\times2}$ and $s_t>0$.
Thus, the allowed transformations are precisely compositions of rotations, reflections, and uniform scalings.

At time $t$, robot $r_t$ executes the following Look--Compute--Move cycle.
\begin{itemize}
\item \textbf{Look:} The robot observes its surroundings and obtains a snapshot $S_t$.
The snapshot is the multiset of the positions of all $n$ robots, including $r_t$ itself, expressed in its local coordinate system with $r_t$ at the origin $(0,0)$:
\[
  S_t=\mset{L_t(p_t(r_i)-p_t(r_t))\mid i\in[0,n-1]}.
\]
\item \textbf{Compute:} The robot computes a destination $x_t\in\R^2$ from $S_t$.
All correct robots use the same function $\calA$ and set $x_t=\calA(S_t)$.
A Byzantine robot may choose $x_t$ arbitrarily.
\item \textbf{Move:} The robot moves to $y_t=p_t(r_t)+L_t^{-1}x_t$.
The move is completed before $r_{t+1}$ is activated at time $t+1$, so $p_{t+1}(r_t)=y_t$.
Since only $r_t$ is activated at time $t$, all other robots remain in place: $p_{t+1}(r_j)=p_t(r_j)$ for every $r_j\ne r_t$.
\end{itemize}

An \emph{algorithm} in this model is the common function $\calA$ that maps a snapshot to a destination.
In particular, correct robots always compute the same destination when given the same snapshot.

\begin{definition}[Gathering]
Let $n\ge2$ and $0\le f<n$ be integers.
An algorithm $\calA$ \emph{solves the $(n,f)$-gathering problem} if, for every initial configuration, every choice of the $f$ Byzantine robots, every behavior of those robots, and every choice of the coordinate transformations $L_t$, all correct robots executing $\calA$ gather at a single point in finite time and remain there forever.
\end{definition}

\begin{definition}[Resilience]
For an integer $n\ge2$, let $\psi(n,\calA)$ be the largest integer $k\in\{0,\dots,n-1\}$ such that $\calA$ solves the $(n,f)$-gathering problem for every integer $0\le f\le k$.
If no such $k$ exists, set $\psi(n,\calA)=-1$.
We define $\psi(n)=\max_{\calA}\psi(n,\calA)$, where the maximum ranges over all possible algorithms.
\end{definition}

\section{Lower bound on resilience}
\label{sec:lower-bound}
In this section, we prove that $\psi(n)\ge\lfloor\sqrt{n}/2\rfloor$ by giving an algorithm that solves the $(n,f)$-gathering problem whenever $f\le\lfloor\sqrt{n}/2\rfloor$.
Under \tomax, an activated robot moves to a point of maximum multiplicity in its snapshot; if several such points exist, it chooses any one other than its current position.
Thus, a robot remains stationary if and only if its current position is the unique point of maximum multiplicity.

\begin{algorithm}
\caption{\tomax}
\Input{Snapshot $S$}
Let $M=\max_{p\in S}m_S(p)$\;
Let $C=\{p\in S\mid m_S(p)=M\}$\;
\If{$|C|=1$}{
  Let $p_{\max}$ be the unique point in $C$\;
}\Else{
  Let $p_{\max}$ be an arbitrary point in $C\setminus\{(0,0)\}$\\\tcp*{
    Ties are resolved by any fixed deterministic rule.
  }
}
Return $p_{\max}$\;
\end{algorithm}

\begin{lemma}
\label{lem:finite-movers-same-point}
In any execution of \textnormal{\tomax}, if two correct robots $r_i$ and $r_j$ each move only finitely many times, then their final positions coincide.
\end{lemma}
\begin{proof}
Consider a suffix of the execution in which both robots remain stationary, and suppose that their positions $p_i$ and $p_j$ are distinct.
After exchanging the roles of the two robots if necessary, assume $i<j$ and let $S=\{r_{i+1},\dots,r_{j-1}\}$.
At any time, let $s_i$ and $s_j$ be the numbers of robots in $S$ located at $p_i$ and $p_j$, respectively, and define the potential $\phi=s_j-s_i$.

Under \tomax, a correct robot can remain stationary only if its current position is the unique point of maximum multiplicity.
Consequently, the difference between the multiplicities of $p_j$ and $p_i$, counting all robots, is negative immediately before $r_i$ is activated and positive immediately before the next activation of $r_j$.
Only robots in $S$ can change position between these two activations, so $\phi$ increases by at least $2$ over this interval.
No robot in $S$ is activated between that activation of $r_j$ and the next activation of $r_i$, so $\phi$ remains unchanged during the rest of the round.
Thus, the values of $\phi$ immediately before successive activations of $r_i$ increase by at least $2$ per round, contradicting $-|S|\le\phi\le|S|$.
\end{proof}

\begin{theorem}
\label{thm:sqrt-lower-bound}
For every $n\ge2$,
$\psi(n,\textnormal{\tomax})\ge
  \left\lfloor\frac{\sqrt{n}}{2}\right\rfloor$.
\end{theorem}
\begin{proof}
Fix an integer $f$ with $0\le f\le \lfloor\sqrt{n}/2\rfloor$,
so that $n\ge 4f^2$.
If $f=0$, all robots gather within the first round, so assume $f\ge1$.
Fix an arbitrary execution.
Let $H$ be the set of correct robots that move only finitely many times, and let $U$ be the set of correct robots that move infinitely often.
Write $h=|H|$ and $u=|U|$, so that $h+u=n-f$.
By Lemma~\ref{lem:finite-movers-same-point}, it suffices to prove $u=0$.

Partition the cyclic activation order into maximal consecutive blocks $A_0,\dots,A_{k-1}$ of correct robots, each followed by a maximal consecutive block $B_i$ of Byzantine robots.
Since every $B_i$ is nonempty, $k\le f$.
Immediately after a correct robot is activated, its position is the unique point of maximum multiplicity.
Hence, after the first full activation of each block $A_i$, all robots in that block are co-located.
In every subsequent round, either all robots in $A_i$ move to the same point or all remain stationary.
It follows that each $A_i$ is contained entirely in either $H$ or $U$.
Write $a_i=|A_i|$.

Suppose $u>0$.
We first claim that $a_i\le2f$ whenever $A_i\subseteq U$.
Indeed, suppose $A_i\subseteq U$ and $a_i\ge2f+1$.
Consider a move of $A_i$ after its first full activation,
when its robots have already become co-located.
Let $p$ be its destination.
Immediately after the block has moved, the multiplicity of $p$ exceeds that of every other point by at least $a_i$.
As long as $p$ is uniquely maximal, correct activations cannot decrease this gap, while each Byzantine activation decreases it by at most $2$.
Since at most $f$ Byzantine activations occur before the next activation of $A_i$, the gap remains at least $a_i-2f>0$ throughout this interval.
Every correct robot outside $A_i$ is activated during this interval and moves to $p$ or remains there.
Thus, all correct robots gather at $p$.
Since $n-f\ge a_i>f$, they remain there forever, contradicting $A_i\subseteq U$.

Using $k\le f$, we obtain
\[
  u\le2f^2,\qquad
  h=n-f-u\ge2f^2-f>0.
\]
Thus, at least one block of correct robots is contained in $H$, leaving at most $f-1$ blocks contained in $U$.
It follows that $u\le2f(f-1)$ and hence
\[
  h=n-f-u\ge n-2f^2+f\ge\frac n2+f>\frac n2.
\]
By Lemma~\ref{lem:finite-movers-same-point}, all robots in $H$ eventually remain stationary at a common point $q$.
Since these robots alone constitute a strict majority of all robots, $q$ is thereafter always the unique point of maximum multiplicity.
Every remaining correct robot therefore moves to $q$ at its next activation and remains there, contradicting $u>0$.
\end{proof}

\begin{corollary}
  $\psi(n) = \Omega(\sqrt{n})$.
\end{corollary}

\section{Upper bound on resilience}
\label{sec:upper-bound}

\subsection{General upper bound}

A corresponding threshold follows from \cite[Lemmas~17 and~18]{DPR20} by setting $f=\lceil n/2\rceil$ and $k=1$.
Since their specification additionally requires the gathering point to be the unique maximum-multiplicity point, we give a self-contained proof for the gathering specification used here.\footnote{When $n>2f$, gathering all correct robots at one point makes it the unique maximum-multiplicity point.
When $n\le2f$, however, Byzantine robots may form an equally large or larger group elsewhere even after all correct robots have gathered and stopped.
Their impossibility result for a stronger specification therefore does not directly imply ours, so a separate proof is needed.}

\begin{theorem}
\label{thm:general-upper-bound}
For every integer $n\ge2$, $\psi(n)\le\left\lfloor\frac{n-1}{2}\right\rfloor$.
\end{theorem}
\begin{proof}
Fix $n\ge2$, and set $f=\lceil n/2\rceil$ and $g=n-f$.
Suppose, for a contradiction, that some algorithm $\calA$ solves both the $(n,0)$- and $(n,f)$-gathering problems.
Use identity coordinate transformations throughout.
In a fault-free configuration with $g$ robots at one point and $f$ at another, some robot must eventually move.
Since $f\ge g$, its snapshot $S$ immediately before the first move satisfies $m_S((0,0))\ge g$ and $\calA(S)\ne(0,0)$.

Now consider an execution with $f$ Byzantine robots.
Repeatedly keep them stationary until all correct robots meet at a point $q$, then use their next activations to reproduce $S$ relative to $q$.
This is possible as long as the correct robots remain stationary, since they account for $g$ occurrences of the origin in $S$.
If a correct robot moves during this placement, restart the procedure immediately.
Otherwise, keep the Byzantine robots at their assigned positions; the next correct robot observes $S$ and moves, after which the procedure restarts.
Each waiting phase ends by correctness for $(n,f)$, so this forces infinitely many correct moves, contradicting eventual stationarity.
Thus, $\psi(n)\le f-1=\lfloor(n-1)/2\rfloor$.
\end{proof}

\begin{figure}[t]
\centering
\definecolor{tomaxcorrect}{HTML}{245F8D}
\definecolor{tomaxbyzantine}{HTML}{B34D16}
\begin{tikzpicture}[
  x=1cm,y=1cm,
  font=\footnotesize,
  site/.style={circle,draw=black!65,fill=white,line width=.45pt,
    minimum size=8mm,inner sep=1pt,font=\small},
  maximum/.style={site,draw=black,line width=1.3pt},
  triangle/.style={draw=black!25,line width=.45pt},
  move/.style={-{Stealth[length=2mm,width=1.4mm]},line width=1pt},
  correct move/.style={move,draw=tomaxcorrect},
  byzantine move/.style={move,draw=tomaxbyzantine},
  occupants/.style={align=center,anchor=north,inner sep=1pt},
  heading/.style={font=\small,anchor=south}
]
\node[anchor=south] at (6.9,3.05)
  {\textcolor{tomaxcorrect}{Correct: $U$ ($k$ robots), $H$ ($f$ robots)}
   \qquad
   \textcolor{tomaxbyzantine}{Byzantine: $b_0$, $B=\{b_1,\ldots,b_{f-1}\}$}};

\begin{scope}
  \node[heading] at (1.2,2.68) {(a) Initial};
  \draw[triangle] (0,0) -- (2.4,0) -- (1.2,2.07846) -- cycle;
  \node[site] (pa) at (0,0) {$k+1$};
  \node[site] (qa) at (2.4,0) {$f-1$};
  \node[maximum] at (1.2,2.07846) {$f$};
  \node[occupants] at (0,-.49) {$p$\\\textcolor{tomaxcorrect}{$U$}, \textcolor{tomaxbyzantine}{$b_0$}};
  \node[occupants] at (2.4,-.49) {$q$\\\textcolor{tomaxbyzantine}{$B$}};
  \node[anchor=west,inner sep=2pt] at (1.65,2.07846) {$z$: \textcolor{tomaxcorrect}{$H$}};
  \draw[byzantine move] (pa) to[bend left=24]
    node[above,text=tomaxbyzantine] {$b_0$} (qa);
\end{scope}

\begin{scope}[xshift=3.8cm]
  \node[heading] at (1.2,2.68) {(b) After $b_0$ moves};
  \draw[triangle] (0,0) -- (2.4,0) -- (1.2,2.07846) -- cycle;
  \draw[densely dashed,draw=black!55,line width=.6pt]
    (-.22,-.127) -- (2.75,1.588);
  \node[site] (pb) at (0,0) {$k$};
  \node[maximum] (qb) at (2.4,0) {$f$};
  \node[maximum] at (1.2,2.07846) {$f$};
  \node[occupants] at (0,-.49) {$p$\\\textcolor{tomaxcorrect}{$U$}};
  \node[occupants] at (2.4,-.49) {$q$\\\textcolor{tomaxbyzantine}{$b_0,B$}};
  \node[anchor=west,inner sep=2pt] at (1.65,2.07846) {$z$: \textcolor{tomaxcorrect}{$H$}};
  \draw[correct move] (pb) to[bend left=24]
    node[above,text=tomaxcorrect,fill=white,inner sep=1pt] {$U$} (qb);
\end{scope}

\begin{scope}[xshift=7.6cm]
  \node[heading] at (1.2,2.68) {(c) After $U$ moves};
  \draw[triangle] (0,0) -- (2.4,0) -- (1.2,2.07846) -- cycle;
  \node[site] (pc) at (0,0) {$0$};
  \node[maximum] (qc) at (2.4,0) {$f+k$};
  \node[site] at (1.2,2.07846) {$f$};
  \node[occupants] at (0,-.49) {$p$\\empty};
  \node[occupants] at (2.4,-.49) {$q$\\\textcolor{tomaxcorrect}{$U$}, \textcolor{tomaxbyzantine}{$b_0,B$}};
  \node[anchor=west,inner sep=2pt] at (1.65,2.07846) {$z$: \textcolor{tomaxcorrect}{$H$}};
  \draw[byzantine move] (qc) to[bend right=24]
    node[above,text=tomaxbyzantine] {$B$} (pc);
\end{scope}

\begin{scope}[xshift=11.4cm]
  \node[heading] at (1.2,2.68) {(d) After $B$ moves};
  \draw[triangle] (0,0) -- (2.4,0) -- (1.2,2.07846) -- cycle;
  \node[site] at (0,0) {$f-1$};
  \node[site] at (2.4,0) {$k+1$};
  \node[maximum] at (1.2,2.07846) {$f$};
  \node[occupants] at (0,-.49) {$p$\\\textcolor{tomaxbyzantine}{$B$}};
  \node[occupants] at (2.4,-.49) {$q$\\\textcolor{tomaxcorrect}{$U$}, \textcolor{tomaxbyzantine}{$b_0$}};
  \node[anchor=west,inner sep=2pt] at (1.65,2.07846) {$z$: \textcolor{tomaxcorrect}{$H$}};
  \node[text=tomaxcorrect,align=center] at (1.2,.66) {$H$ stays\\($k+1<f$)};
\end{scope}

\foreach \offset in {0,3.8,7.6}{
  \draw[-{Stealth[length=1.6mm]},draw=black!50,line width=.6pt]
    (\offset+2.93,1.05) -- (\offset+3.4,1.05);
}
\draw[-{Stealth[length=1.8mm]},draw=black!60,line width=.7pt]
  (12.6,-1.16) -- (12.6,-1.46) -- (1.2,-1.46) -- (1.2,-1.16);
\node[fill=white,inner xsep=5pt,inner ysep=1pt] at (6.9,-1.46)
  {Repeat with $p$ and $q$ exchanged};
\end{tikzpicture}
\caption{One round of the repeating execution in Theorem~\ref{thm:tomax-upper-bound}.
Numbers inside circles are total multiplicities; thick circles mark points of maximum multiplicity.
Arrows show the moves leading to the next panel.
The dashed line in~(b) is the reflection axis fixing $p$ and exchanging $q,z$.
In~(d), $H$ stays at $z$, and the initial configuration is recovered with $p,q$ exchanged.}
\label{fig:tomax-upper-bound}
\end{figure}
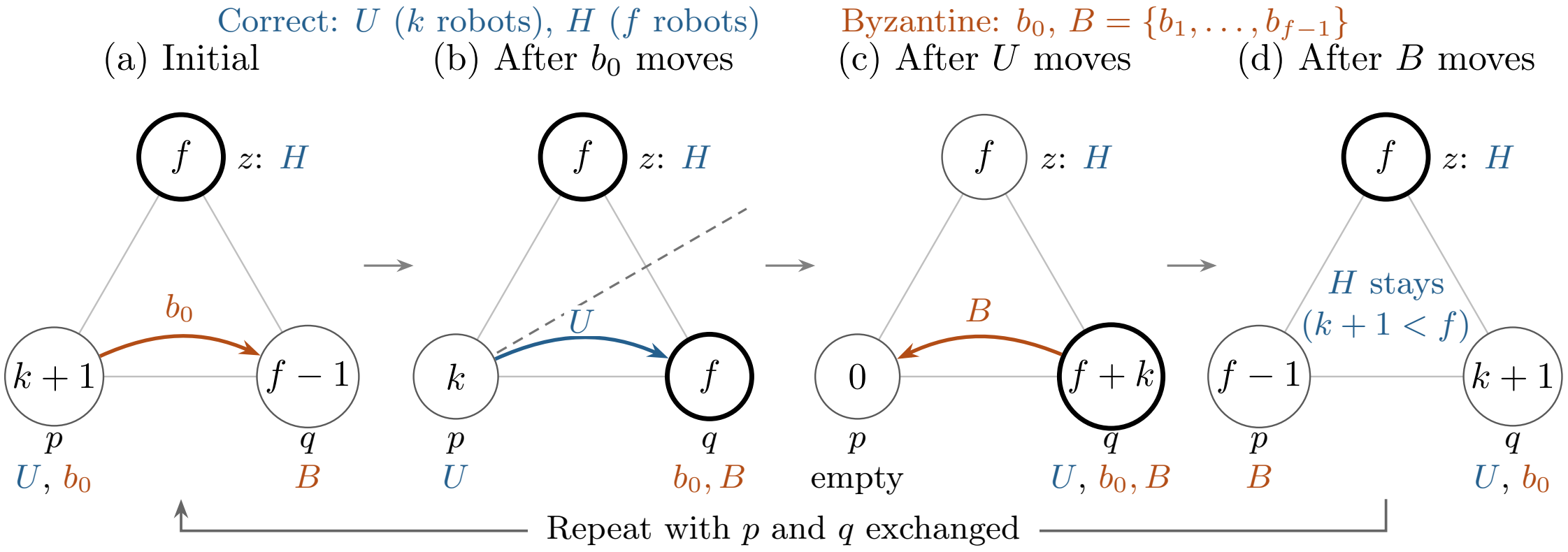

\subsection{Upper bound for \tomax}

\begin{theorem}
\label{thm:tomax-upper-bound}
For every integer $n\ge2$,
$\psi(n,\textnormal{\tomax})\le
  \left\lfloor\frac{n+1}{3}\right\rfloor$.
\end{theorem}
\begin{proof}
Set $f=\lceil(n+2)/3\rceil$.
If $n\le2f$, Theorem~\ref{thm:general-upper-bound} gives the desired bound $f-1$.
Otherwise, set $k=n-2f$, so $1\le k\le f-2$.
Let $p,q,z$ be the vertices of an equilateral triangle.
Partition the correct robots into blocks $U$ and $H$ of sizes $k$ and $f$, respectively, and label the Byzantine robots $b_0,\dots,b_{f-1}$.
Arrange them in the round-robin order
\[
  b_0,\ U,\ b_1,\dots,b_{f-1},\ H.
\]
Initially, place $U$ and $b_0$ at $p$, the remaining Byzantine robots at $q$, and $H$ at $z$.
Figure~\ref{fig:tomax-upper-bound} illustrates one round of the construction.

First, $b_0$ moves to $q$, giving multiplicities $(k,f,f)$ at $(p,q,z)$.
Consider two local coordinate systems related by reflection about the line through $p$ and the midpoint of $qz$.
This reflection fixes $p$ and exchanges the equally populated points $q$ and $z$, so the first robot in $U$ sees the same snapshot in both systems.
Its deterministic rule therefore returns the same local destination, while the corresponding physical destinations are exchanged between $q$ and $z$.
Thus, the adversary can choose a coordinate system that makes this robot move to $q$, regardless of the deterministic tie-breaking rule.
Then $q$ is uniquely maximal, so the rest of $U$ follows.
Next, $b_1,\dots,b_{f-1}$ move to $p$, giving multiplicities $(f-1,k+1,f)$.
Since $k+1<f$, all robots in $H$ remain at $z$.
The resulting configuration is the initial one with $p$ and $q$ exchanged, so this execution repeats forever with $U$ alternating between them.
Hence \tomax{} fails for $(n,f)$, proving $\psi(n,\textnormal{\tomax})\le f-1=\lfloor(n+1)/3\rfloor$.

\end{proof}

\section{Conclusion}
\label{sec:conclusion}

We showed that \tomax{} solves deterministic gathering under RR scheduling whenever $f\le\lfloor\sqrt{n}/2\rfloor$, assuming oblivious robots, rigid movement, and strong multiplicity detection.
This extends the previously known resilience guarantee from a single Byzantine robot to a number that grows with the system size.

A substantial gap remains between our square-root lower bound and the linear upper bounds, both for \tomax{} and for the optimal resilience $\psi(n)$.
In particular, whether RR can tolerate a constant fraction of Byzantine robots remains open.

\paragraph{AI disclosure.}
We used Codex (ChatGPT-6 Astra) to help check proofs, improve the exposition, and verify statements about related work.
Codex also generated an initial version of Figure~\ref{fig:tomax-upper-bound}.
The authors have verified all content, including the references, and take full responsibility for its correctness and originality.

\bibliographystyle{alpha}
\bibliography{byzRobots}

\clearpage

\end{document}